\documentclass[prd,twocolumn,showpacs,showkeys,preprintnumbers,amsmath,amssymb]{revtex4}
\usepackage{graphicx}
\usepackage{dcolumn}
\usepackage{bm}

\newcommand{\etal}{{\it et al.}, }

\begin{document}

\title[Cosmogenic $^{11}$C production]{Cosmogenic $^{11}$C production \\
and sensitivity of organic scintillator detectors \\
to pep and CNO neutrinos}

\author{Cristiano Galbiati}
\email[Correspinding author.  E-mail address:]{galbiati@princeton.edu}
\author{Andrea Pocar}
\altaffiliation{now at Physics Department, Stanford University, Stanford, California 94305}
\affiliation{Physics Department, Princeton University, Princeton, New Jersey 08544}
\author{Davide Franco}
\altaffiliation{also at Max-Planck-Institut f\"ur Kernphysik, Heidelberg D 69117, Germany}
\affiliation{Dipartimento di Fisica, Universit\`a degli Studi di Milano, I-20133 Milano, Italy}
\author{Aldo Ianni}
\affiliation{Laboratori Nazionali del Gran Sasso, I-67010 Assergi, Italy}
\author{Laura Cadonati}
\affiliation{Physics Department, Massachusetts Institute of Technology, Cambridge, Massachusetts 02139}
\author{Stefan Sch\"onert}
\affiliation{Max-Planck-Institut f\"ur Kernphysik, D-69117 Heidelberg, Germany}

\date{\today}

\begin{abstract}
Several possible background sources determine the detectability of pep and CNO solar neutrinos in organic liquid scintillator detectors.  Among such sources, the cosmogenic $^{11}$C nuclide plays a central role.  $^{11}$C is produced underground in reactions induced by the residual cosmic muon flux.  Experimental data available for the effective cross section for $^{11}$C by muons indicate that $^{11}$C will be the dominant source of background for the observation of pep and CNO neutrinos.  $^{11}$C decays are expected to total a rate 2.5 (20) times higher than the combined rate of pep and CNO neutrinos in Borexino (KamLAND) in the energy window preferred for the pep measurement, between 0.8 and 1.3~MeV.

This study examines the production mechanism of $^{11}$C by muon-induced showers in organic liquid scintillators with a novel approach:  for the first time, we perform a detailed {\em ab initio} calculation of the production of a cosmogenic nuclide, $^{11}$C, taking into consideration all relevant production channels.  Results of the calculation are compared with the effective cross sections measured by target experiments in muon beams.

This paper also discusses a technique for reduction of background from $^{11}$C in organic liquid scintillator detectors, which allows to identify on a one-by-one basis and remove from the data set a large fraction of $^{11}$C decays. The background reduction technique hinges on an idea proposed by Martin Deutsch, who suggested that a neutron must be ejected in every interaction producing a $^{11}$C nuclide from $^{12}$C.  $^{11}$C events are tagged by a three-fold coincidence with the parent muon track and the subsequent neutron capture on protons.
\end{abstract}

\pacs{
25.20.-x;
25.30.Mr;
26.65.+t;
28.20.Gd;
96.40.Tv;
96.60.-j
}
\keywords{
Muon-induced nuclear reactions;
Photonuclear reactions;
Neutron moderation and diffusion;
Solar neutrinos;
Low background experiments
}
\maketitle

\section{Introduction}
\label{s:introduction}
Observation of solar neutrinos over the past 35 years by seven experiments has offered a unique opportunity to probe particle physics beyond the Standard Model of electroweak and strong interactions and physics of the stellar models. Radiochemical experiments have measured the combined flux of a number of different neutrino sources~\cite{chlorine,gallex,sage,gno}.  The only solar neutrinos targeted by a real time measurement thus far have been the $^8$B neutrinos above a detection threshold of about 5~MeV~\cite{kamiokande,superkamiokande,sno}.  Plans are in place to measure the sub-MeV  neutrino spectrum with organic liquid scintillator-based detectors, focusing on the presumably abundant $^7$Be neutrinos, with Borexino~\cite{borex}, KamLAND~\cite{kamland}, and a possible detector at SNOLab~\cite{mchen}.  Such detectors also have the potential to probe the intermediate energy region, searching for the less abundant pep and CNO neutrinos.
 
The possibility of detecting pep neutrinos in organic liquid scintillator based detectors is particularly intriguing.  It was recently pointed out~\cite{roadmap} that a measurement of the flux of pep solar neutrinos would yield essentially equivalent information about neutrino oscillation parameters and the other solar neutrino fluxes as a measurement of pp solar neutrinos at a comparable level of experimental uncertainty.  Moreover, given the low theoretical uncertainty on the pep solar neutrinos flux, its measurement could allow investigating the matter-vacuum transition region for solar neutrino oscillations~\cite{roadmap}.  We recall that a matter-vacuum transition of solar neutrino oscillation is expected in the region between 2 and 3~MeV for the MSW-LMA solution of the solar neutrino problem, and its observation would provide a further stringent test of the MSW-LMA solution.  Due to this transition, the survival probability - i.e. the probability that electron neutrinos emitted by the sun have not oscillated into other neutrino species when they arrive on earth - for pep neutrinos is about a factor of two larger than the one measured at higher energy by the SNO experiment for $^8$B neutrinos~\cite{sno}.

The interaction rates for pep and CNO solar neutrinos in organic scintillator predicted by the BP04 version of the Standard Solar Model by Bahcall and Pinsonneault~\cite{bp04}, in the currently preferred MSW-LMA solution for the solar neutrino problem~\cite{roadmap}, are 2.1 and 6.6 events per day in 100 metric tons of liquid scintillator respectively.  When data recently released by the LUNA collaboration for the $^{14}$N + p fusion cross section~\cite{luna} are taken into consideration, the signal rate from CNO neutrinos is significantly decreased to an expected 3.5 events per day in 100 tons.

\begin{figure}[t]
\includegraphics[width=3.4in]{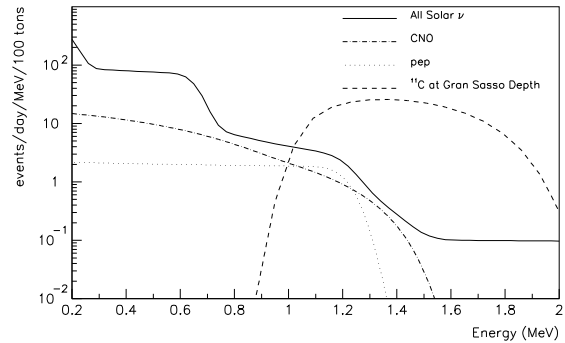}
\caption{\label{f:spectrum} Energy spectrum of electrons scattered by pep (dotted line) and CNO neutrinos (dash-dotted line), for the MSW-LMA solution of the solar neutrino problem.  The total spectrum for all solar neutrinos (including pp, $^7$Be, and $^8$B neutrinos, not shown separately) is also shown (continuous line).  Neutrino fluxes for the spectra shown are from the BP04 version of the Standard Solar Model (see reference~\cite{roadmap}).  The background signal expected from cosmogenic $^{11}$C at Gran Sasso depth is superimposed (dashed line).  For all the neutrino and the $^{11}$C spectra, we assume that the energy resolution of the detector is 5\% at 1~MeV and varies with the energy E as $\rm 1/\sqrt{E}$.}
\end{figure}

The pep neutrino energy spectrum is distinctive, with a single 1.44~MeV monochromatic line.  The energy spectrum for electrons scattered in $\nu$-e interactions presents a characteristic Compton-like edge at 1.22~MeV.  Figure~\ref{f:spectrum} shows the expected (MSW-LMA) spectrum of scattered electrons from different neutrino sources in the energy range of interest for this study.  We will focus our attention on the detection and measurement of the pep neutrino line.  For this purpose we will set an observation window for the recoil electron between 0.8 and 1.3~MeV: for sake of simplicity, all event rates cited in the following will refer to this energy window, unless otherwise noted.  The expected signal rate S in said window for pep and CNO neutrinos combined is 2.0 events per day in 100 tons (1.5 events per day per 100 tons when using the most recent results from LUNA).

There are three fundamental prerequisites for a succesful measurement of the pep and CNO solar neutrinos in an organic liquid scintillator detector.

First, the internal background from long-lived radioactive sources must be carefully controlled.  A $^{238}$U and $^{232}$Th contamination at the 10$^{-17}$ g/g level, coupled with a $^{\rm nat}$K contamination of 10$^{-15}$ g/g would produce  0.6 background events per day in 100 tons in the observation window~\cite{borex,cadonati}; this is below the event rate expected for pep neutrinos.  Contamination from long-lived radon daughters out of secular equilibrium with $^{238}$U (in particular from $^{210}$Bi) must also be reduced below one count per day in the window of interest~\cite{ctf, borex, kamland}.

A second prerequisite is a low external gamma-ray background from the construction materials of the detector and from the surrounding rocks.  The general strategy to solve the problem is to use pure buffer materials to screen environmental radioactivity present in underground laboratories~\cite{borex} and extremely low radioactivity construction materials~\cite{borex-rad}.  The typical spherical geometry of liquid scintillator based detectors has a peculiar effect on the spectrum of gamma-induced events reconstructed within a given radius: the spectrum gets harder at smaller radii.  Therefore, in the innermost part of the detector the $\gamma$ ray-induced background in the pep window [0.8-1.3~MeV] is larger than the corresponding background in the $^7$Be window [0.25-0.8~MeV].  This potential problem may be counterbalanced by redefining the fiducial mass for the observation of pep and CNO neutrinos.  For reference, in the 100~tons fiducial mass of Borexino for $^7$Be neutrinos detection the external background in the [0.8-1.3~MeV] window is expected at 1~event/day~\cite{borex,cadonati}, compared with a neutrino signal of 2~events/day.  Reducing the mass to 70~tons would lower the background by a factor 10~\cite{borex,cadonati} while losing only 30\% of the signal.

The third fundamental condition for the observation of pep and CNO neutrinos is a low internal cosmogenic background production.  This topic is the main subject of this paper.  Section~\ref{s:background} is an introduction to the problem of $^{11}$C cosmogenic background in liquid scintillator detectors.  Section~\ref{s:rates} describes the production channels for cosmogenic $^{11}$C and the results of our calculations for the expected production rate in a 100 tons liquid scintillator target.  Section~\ref{s:comparison} offers a comparison of our estimate with another estimates available in the literature for the production of $^{11}$C by direct interaction of a muon with a $^{12}$C nucleus through virtual photons.  Finally, section~\ref{s:reduction} proposes a veto mechanism for $^{11}$C production events, based on the double coincidence with the parent muon and a neutron produced in the reaction.  We draw our conclusions in section~\ref{s:conclusions}.

\section{Cosmogenic background in deep underground detectors}
\label{s:background}

Cosmogenic radioactive nuclides are produced in deep underground detectors in reactions triggered by the residual muon flux.  As originally pointed out in~\cite{olga}, the fundamental scale of the process is given by the neutron production rate: neutrons are an important and ubiquitous by-product of cosmic ray-induced nuclear reactions.

A comprehensive review of the experimental results on neutron production in underground laboratories can be found in~\cite{wang}.  The first measurement at Gran Sasso (depth of 3800 meters of water equivalent, muon flux of 1.2 m$^{-2}$ hr$^{-1}$, average muon energy 320 GeV~\cite{macro}) was performed with the CTF detector~\cite{ctf} in 1995 yielding $(1.5 \pm 0.1)\times 10^{-2} \  {\rm neutrons}/\mu/{\rm m}$ (neutrons per meter of $\mu$ track) in scintillator with a density of 0.88 g/cm$^3$~\cite{borex,deutsch1,galbiati}.  A later measurement by the LVD experiment gave $(1.4 \pm 0.4)\times 10^{-2} \  {\rm neutrons}/\mu/{\rm m}$~\cite{LVD}.  The figure from the CTF experiment translates into a neutron capture rate of 40 events per day in 100 tons of scintillator.

Deutsch suggested~\cite{deutsch1} that the underground production rate for all of the most significant cosmogenic nuclides out of a target mass composed of $^{12}$C and $^1$H could be estimated from the neutron production rate alone.  The list includes $^8$Li, $^9$Li, $^{11}$Be, $^8$B, $^{12}$B, and $^9$C, all with mean lives below one minute and easily taggable with the parent muon.  $^{7}$Be is the cosmogenic radionuclide with the longest mean life, 77 days.  $^{11}$C also poses problems, given its 30~minutes mean life which does not allow identifying its decays by tagging them  with the parent muon alone.  Deutsch hypothized that $^{11}$C would be one of the most likely cosmogenic nuclides by-products of muon-induced cascades, and estimated that $^{11}$C would account for  5\% of the total neutron production rate.  This would correspond to 2 events per day in the 100 fiducial tons of Borexino.

The inclusive cross section for the production of several cosmogenic nuclides in muon-induced cascades was measured by a target experiment on a muon beam at CERN~\cite{hagner}.  The experiment used a liquid scintillator target.  The muon showers were built up in 240~cm of concrete and 200~cm of water used as absorbers and placed in front of the $^{12}$C targets.  The use of two positively charged muon ($\mu^+$) beams of 100 and 190~GeV allowed to extract information about the energy dependence of the inclusive cross sections.
The results in~\cite{hagner} have shown that $^{7}$Be is not among the most likely products of cosmic ray-induced reactions: its production rate at Gran Sasso depth is expected to be less than 0.1 events per day in 100 tons~\cite{hagner}.  A precise measurement of the inclusive  $^{11}$C production rate was also made available.  At the mean energy of muons at Gran Sasso, the production rate is 15 events per day in 100 tons~\cite{hagner} or, equivalently, $52 \times 10^{-4} \  ^{11}{\rm C}/\mu/{\rm m}$.

The rate is different for other locations due to the different muon rates and muon average energies.  Muon rates and expected $^{11}$C production rates in KamLAND, Borexino, and at SNOLab are summarized in tables~\ref{t:muons} and~\ref{t:c11-extr}.

\begin{table}[t]
\caption{\label{t:muons} Depth, residual muon flux, average muon energy, and neutrons capture rate (N) at KamLAND, Borexino, and SNOLab.  Data on muons are from references~\cite{hagner,macro,cribier,kamland,waltham,tagg,kudryavtsev}.  The neutron capture rate for Borexino comes from the CTF experiment~\cite{borex,deutsch1,galbiati} and the one for KamLAND is the rate measured in the detector and reported in reference~\cite{kamland}.  The neutron capture rate for SNO is by extrapolating the CTF data with the scaling law introduced in~\cite{olga}; when the same procedure is applied to calculate the capture rate in KamLAND, the result of 294~cts/d/100~tons is fully consistent with the value measured by KamLAND.}
\begin{ruledtabular}
\begin{tabular}{lcdcd}
	&\multicolumn{1}{c}{Depth}
	&\multicolumn{1}{c}{$\Phi_\mu$}
	&\multicolumn{1}{c}{$\rm < E_\mu>$}
	&\multicolumn{1}{c}{N} \\
	&\multicolumn{1}{c}{[m.w.e.]}
	&\multicolumn{1}{c}{[$\rm \mu/m^2/h$]}
	&\multicolumn{1}{c}{[GeV]}
	&\multicolumn{1}{c}{[cts/d/100  tons]} \\
\hline
KamLAND	&2700	&9.6		&285	&300 \\
Borexino		&3800	&1.2		&320 	&40 \\
@ SNOLab	&6000	&0.012	&350 	&0.43 \\
\end{tabular}
\end{ruledtabular}
\end{table}

\begin{table}[t]
\caption{\label{t:c11-extr} Total expected $^{11}$C decay rate and $^{11}$C-induced raw background rate (B$_0$) in the pep window [0.8-1.3~MeV] at KamLAND, Borexino, and SNOLab (approximately 35\% of $^{11}$C decays produce an event the pep window).  Data for Borexino and KamLAND are from reference~\cite{hagner}, extrapolated from data of target experiment on muon beam at 100 and 190 GeV.  Data for SNOLab are calculated from the muon flux and average energy reported in table~\ref{t:muons} using the same extrapolation method of reference~\cite{hagner}.}
\begin{ruledtabular}
\begin{tabular}{ldcd}
	&\multicolumn{2}{c}{$^{11}$C rate}
	&\multicolumn{1}{c}{B$_0$} \\
	&\multicolumn{1}{c}{[cts/d/100 tons]}
	&\multicolumn{1}{c}{[$\rm 10^{-4} /\mu/{\rm m}$]}
	&\multicolumn{1}{c}{[cts/d/100 tons]} \\
\hline
KamLAND	&107		&48		&37 \\
Borexino		&15			&52		&5.1 \\
@ SNOLab	&0.15		&55		&0.056 \\
\end{tabular}
\end{ruledtabular}
\end{table}

\section{$^{11}$C production in muon-induced showers}
\label{s:rates}

$^{11}$C is a positron emitter with a 0.96~MeV end point; approximately 35\% of its decays produce an event in the pep window [0.8-1.3~MeV].  The raw background rate (B$_0$) from $^{11}$C in the pep window varies with the depth of the location of the experiments.  Values for the raw background for Borexino, KamLAND, and at SNOLab are also reported in table~\ref{t:c11-extr}.  The background at SNOLab is sufficiently low to enable pep and CNO neutrinos observation without need of any cuts on $^{11}$C.

Deutsch pointed out that the only way to create $^{11}$C is to knock a neutron off the $^{12}$C nucleus and suggested to look for a neutron in the final stated of the reaction, emphasizing the possibility of a three-fold coincidence with the parent muon track and the neutron capture on protons in the scintillator to tag the $^{11}$C events on a one-by-one basis~\cite{deutsch1}.  On the other hand, as suggested by Calaprice, there is also the possibility of creating a $^{11}$C while ejecting a deuteron in a (p,d) exchange reaction.  The (p,d) reaction would create an invisible channel for $^{11}$C production since the nuclide produced through such process cannot be tagged with the three-fold coincidence mentioned above.  Similarly, reactions triggered by $\pi$ mesons can also produce decays in invisible channels, as explained later.

Following is a list of the leading reactions that can produce $^{11}$C, together with their cross section and references to specific studies.
For all the reactions with a neutron in the final state, the energy threshold is $\sim20\,$MeV (i.e. the neutron binding energy in $^{12}$C).

\begin{figure}[t]
\includegraphics[width=3.4in]{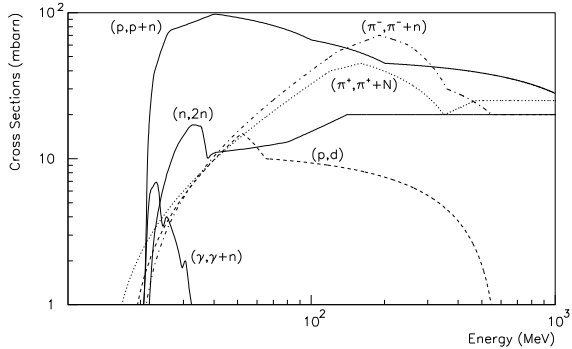}
\caption{\label{f:cross_sections} Cross sections for $^{11}$C production from $^{12}$C as a function of energy.}
\end{figure}

\begin{description}
\item[$\bm ^{12}$C($\bm \gamma$,n)$\bm ^{11}$C]\cite{gamma-n}: the cross section for the $\gamma$-ray induced process peaks at 7~mb around 23~MeV.  The cross section value in the peak region, relatively large for an electromagnetic interaction, is due to the nuclear giant dipole resonance~\cite{goldhaber}.
\item[$\bm ^{12}$C(n,2n)$\bm ^{11}$C]\cite{n-2n}: the cross section has a sharp peak (17~mb) around 33~MeV.  At higher energies, we rely on the set of experimental data by Kim~{\it et~al.}~\cite{n-2n} in the range 40-150~MeV, which is the only available for energies above 40~MeV. Those data are affected by a large experimental error, as high as 40\%.  As pointed out by the authors, their results disagree starkly with theoretical expectations which predict much lower values for the cross section in the same energy range~\cite{endf}.  Also, we assume that at energies higher than 150~MeV the cross section keeps the constant value attained in the range 70-150~MeV.  Uncertainties in the knowledge of the cross section for this process represent the largest systematic error in our {\it ab initio} calculation of the rate of production of cosmogenic $^{11}$C.  We estimate that the systematic error attributable to this source could reach 50\% of the production rate expected from the $^{12}$C(n,2n)$^{11}$C channel.  All the other cross sections are known with a precision better than a few percent.
\item[$\bm ^{12}$C(p,p+n)$\bm ^{11}$C]\cite{p-pn}: the cross section reaches a peak value of 98~mb at 40~MeV, and decreases to a plateau of 30~mb, constant up 1~GeV.
\item[$^{12}$C(p,d)$^{11}$C]\cite{p-d}: the cross section has a threshold of 16~MeV and has been measured at 52 and 65~MeV (15 and 10~mb respectively).  The only measurement available in the range above 100~MeV tells us that the cross section is in the range of a few $\mu$b, and therefore negligible.
\item[$\bm ^{12}$C($\bm \pi^-$,$\bm \pi^-$+n)$\bm ^{11}$C]\cite{pions}: the cross sections exhibits a broad peak centered around the (3,3) resonance for the pion-nucleon quasi-elastic interaction (see Dropesky {\it et al.}~\cite{pions}) with a value of 70~mb at 190~MeV.  Data are available up to 550~MeV, and show that the cross section reaches a plateau above 400~MeV.  We assume that the cross section keeps a constant value at higher energies.
\item[$^{12}$C($\bm \pi^+$,$\bm \pi$+N)$^{11}$C]\cite{pions}: the cross section exhibits a broad resonance peak in the same region, reaching 45~mb around 160~MeV.   Data are available up to 470~MeV, and show that the cross section reaches a plateau above 350~MeV.  We assume that the cross section keeps a constant value at higher energies.  Contrarily to $\pi^-$, positive mesons do not necessarily produce a neutron in the final state: the N in the final case stands for nucleon and can be either a proton or a neutron.  Due to the possibile charge exhcange occurring in the strong meson-nucleon interaction, the fragments in the final state can be either ($\pi^+$+n) or ($\pi^0$+p), the latter having a threshold of 13~MeV.  No data are available on the relative composition of the final state, but it is expected from theoretical considerations (see Chivers {\it et al.} in~\cite{pions}) that the frequency of the invisible channel, with a proton in the final state, should account for 2/3~of all the $\pi^+$-induced reactions.  We rely on this assumption in our calculations.
\item[$^{12}$C(e,e+n)$^{11}$C]\cite{electrons}: The direct interaction of electrons and positrons with a nucleus, through a virtual photon, is expected to have a small cross section, of the order of $\alpha\,(=1/137)$ times the peak value of the cross section for real photons.  The measured value of the cross section at 30~MeV is 15~$\mu$b.  At higher energies, the cross section can be calculated by using the von~Weizs\"acker approximation~\cite{weizsacker}:
\begin{equation}
\label{e:virtual-ph1}
\rm \sigma_e = \int N(\nu) \sigma_\gamma(\nu) d\nu
\end{equation}
where $\rm \sigma_e$ is the cross section for the $^{11}$C production induced by electrons, $\sigma_\gamma$ is the cross section for $^{11}$C production by real photons, $\nu$ is the energy of virtual photons and N is the number of virtual photons.  In case of high energy, ultra-relativistic charged particles inducing nuclear reactions with low momentum transfer, the number of virtual photons can be approximated by~\cite{virtual-photons}:
\begin{equation}
\label{e:virtual-ph2}
\rm N(\nu) = (\alpha/\pi\nu) \left[  2 \ln{(E/m)}- 1 \right]
\end{equation}
where E is the energy of the charged particle and m is its mass.  The cross section for the production induced by electrons is then given by:
\begin{equation}
\label{e:virtual-ph3}
\rm \sigma_e = (\alpha/\pi) \left[  2 \ln{(E/m)}- 1 \right] \sigma_{-1}
\end{equation}
where $\sigma_{-1} = \int d\nu \sigma_\gamma(\nu)/\nu$ is the inverse energy weighted moment of the photodisintegration cross section $\sigma_\gamma$.  Using the photodisintegration cross sections from references~\cite{gamma-n}, we calculated a value of 1.74~mb for $\sigma_{-1}$.  With this value, the cross sections for the production of $^{11}$C by electrons of 100~MeV, 1~GeV, and 10~GeV are 39~$\mu$b, 57~$\mu$b, and 76~$\mu$b  respectively.  The values of the cross sections at intermediate energies are interpolated from the above values.
\item[$\bm ^{12}$C($\bm \mu$,$\bm \mu$+n)$\bm ^{11}$C] The direct interaction of muons with a nucleus, through a virtual photon, is usually referred to as ``muon spallation".  The cross section for the process can be calculated using the same procedure detailed above for electrons.  The result is a cross section of 58~$\mu$b for muons at 320~GeV.  In the range 100-350~GeV the cross sections have values very close to the one just quoted, given the logarithmic dependence of the number of virtual photons N upon the energy E, as shown in equation~\ref{e:virtual-ph2}.
\end{description}

The cross sections for $^{11}$C production by photons and hadrons, compiled using the procedure detailed above, are shown in figure~\ref{f:cross_sections}.  

We performed a full simulation of muon-induced showers with the particle transport code FLUKA~\cite{fluka}.  The FLUKA code has been used by Wang {\it et al.}~\cite{wang} to calculate the production rate of neutrons by muons in liquid scintillator at several depths, and has been found to reproduce experimental results very well.  Recently, FLUKA has been used by Kudryavtsev {\it et al.}~\cite{kudryavtsev} to calculate the distance between the parent muon track and the point of capture on protons of neutrons produced in scintillator by muon-induced cascades at Gran Sasso depth, and results were found to be in agreement with the experimental data from the LVD experiment~\cite{LVD}.

\begin{figure}[t!]
\includegraphics[width=3.4in]{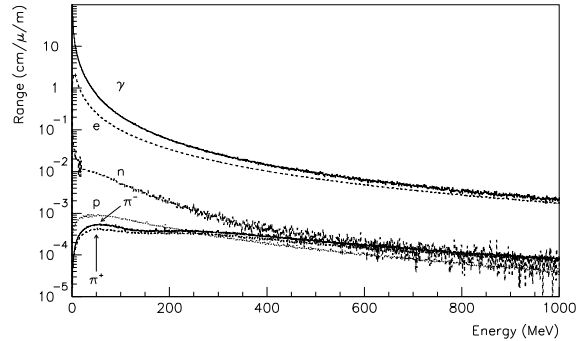}
\caption{\label{f:secondaries_range} Cumulative range of secondaries generated in showers induced by negatively charged muons at 320~GeV.  Results are quoted in cm of range per meter of $\mu$  track.}
\end{figure}

We used FLUKA to calculate production rates and ranges of all the prominent secondaries, i.e. protons, neutrons, $\pi$ mesons, and $\gamma$-rays.  We simulated showers originating from negatively charges muons ($\mu^-$) at 100 and 190 GeV (the energies of the muon beams for the experiement described in~\cite{hagner}), at 285 GeV (average energy at Kamioka), at 320 GeV (average energy at Gran Sasso), and at 350 GeV (average energy at SNOLab).  The target material in the simulation was the solvent of the liquid scintillator for Borexino, trimethylbenzene (C$_9$H$_{12}$), with  density 0.88 g/cm$^3$ (incidentally, this makes up 20\% of the solvent used in KamLAND~\cite{kamland}).  Results should not vary greatly with other organic solvents, given that typical values of density and mass ratio between carbon and hydrogen are close to the values of trimethylbenzene.  We tracked muons for 100 meters, and for each of the prominent secondaries we calculated the cumulative range of the particles as a function of the particle energy with a 10 GeV cutoff.  The results for secondary particles with energy below 1 GeV in showers induced by negatively charged muons at 320~GeV are shown in figure~\ref{f:secondaries_range}.

We then turned to the computation of the $^{11}$C production rate from each one of the interactions listed above, for each of the energies of the muons taken into consideration.  The production rate has been calculated by taking the energy convolution, for each of the possible interactions, of the cross sections with the range of the secondary particles responsible for inducing that particular interaction.  Results are summarized in table~\ref{t:prod-rates}.  The production rate in the invisible channels rate has been calculated by adding the rate from the $^{12}$C(p,d)$^{11}$C reaction to 2/3~of the rate from the $^{12}$C($\pi^+$,$\pi$+N)$^{11}$C reaction.

The error quoted in table~\ref{t:prod-rates} accounts for the systematic error, coming from two main sources.  The most important source is the uncertainty in the knowledge of the cross section for the process $^{12}$C(n,2n)$^{11}$C.  Our best estimate for the systematic error is 50\% of the production rate in such channel, which is between 5\% and 7\% of the total production rate depending on the energy.  We also make the conservative assumption that the systematic error associated with the cross sections used by FLUKA to calculate the range of the secondaries accounts for 5\% of the total production rate.  We combine the two as independent sources of error.  The statistical error associated with our Monte Carlo calculation is 0.6\% and is negligible with respect to the systematic error.

\begin{table}[t!]
\caption{\label{t:prod-rates} Production rates for $^{11}$C in muon induced showers.  The calculated total production rates are compared with the experimental values available at 100 and 190 GeV from Ref.~\cite{hagner}, and with the extrapolated values at the mean muon energy for KamLAND and Borexino (also from Ref.~\cite{hagner}) and at SNOLab.
The procedures used to determine the expected rate from the invisible channels and the systematic error affecting our calculation are outlined in the text.}
\begin{ruledtabular}
\begin{tabular}{lddddd}
	E$_\mu$	[GeV]
	&\multicolumn{1}{c}{100}
	&\multicolumn{1}{c}{190}
	&\multicolumn{1}{c}{285}
	&\multicolumn{1}{c}{320}
	&\multicolumn{1}{c}{350} \\
\hline\hline
						&\multicolumn{5}{c}{Rate}\\
\multicolumn{1}{c}{Process}	&\multicolumn{5}{c}{[$10^{-4}/\mu/{\rm m}$]}\\
\cline{2-6}
$^{12}$C(p,p+n)$^{11}$C			
	&1.8		&3.2		&4.9		&5.5		&5.6 \\
$^{12}$C(p,d)$^{11}$C
	&0.2		&0.4		&0.5		&0.6		&0.6 \\
$^{12}$C($\gamma$,n)$^{11}$C		
	&19.3	&26.3	&33.3	&35.6	&37.4 \\
$^{12}$C(n,2n)$^{11}$C
	&2.6		&4.7		&7.0		&8.0		&8.2 \\
$^{12}$C($\pi^+$,$\pi$+N)$^{11}$C	
	&1.0		&1.8		&2.8		&3.2		&3.3 \\
$^{12}$C($\pi^-$,$\pi^-$+n)$^{11}$C
	&1.3		&2.3		&3.6		&4.1		&4.2 \\
$^{12}$C(e,e+n)$^{11}$C
	&0.2		&0.3		&0.4		&0.4		&0.4 \\
$^{12}$C($\mu$,$\mu$+n)$^{11}$C
	&2.0		&2.3		&2.4		&2.4		&2.4 \\
\hline
Invisible channels
	&0.9		&1.6		&2.4		&2.7		&2.8 \\
\hline
Total
	&28.3	&41.3	&54.8	&59.9	&62.2 \\
1$\sigma$ systematic
	&1.9		&3.1		&4.4		&5.0		&5.2 \\
\hline
Measured
	&22.9	&36.0	&		&		& \\
1$\sigma$ experimental
	&1.8		&2.3		&		&		& \\
Extrapolated
	&		&		&47.8	&51.8	&55.1 \\
\end{tabular}
\end{ruledtabular}
\end{table}

The total calculated rate is systematically 20\% higher than the measured rates on a beam at 100 and 190~GeV, even though the values are still within twice the combined experimental and systematic uncertainties.  A possible explanation for the systematic discrepancy is the difference in the geometry between the experiment and the simulation (in the experiment, the muon shower is built up in 240~cm of concrete and 200~cm of water placed in front of the scintillator target; in our simulation we consider a bulk scintillator volume).  The calculated rates at 285, 320, and 350~GeV, although systematically higher than the extrapolated values, are still in good agreement with them when considering the systematic error.

We also calculated the production rates for positively charged muons ($\mu^+$) at 320 GeV.  The difference in the production of $^{11}$C, for all of the channels considered, by $\mu^+$ and $\mu^-$ at 320 GeV is within the statistical error of our Monte Carlo calculation.  We conclude that the dependence of the inclusive cross section for $^{11}$C production on the sign of the charge of the muons is negligible within the scope of the study presented in this paper.

We took into account also the possibility of producing $^{11}$C nuclides from the target nuclide $^{13}$C, which has a natural isotopic abundance of 1.1\%~\cite{isotopes}.
The contributions of channels such as $^{13}$C($\gamma$,2n)$^{11}$C~\cite{13c-gamma}, $^{13}$C($\pi^+$,d)$^{11}$C~\cite{13c-pi}, and $^{13}$C(p,t)$^{11}$C~\cite{13c-t}, are negligible with respect to the corresponding rates for processes with the same incident particles on the target nuclide $^{12}$C, owing to smaller cross sections and to the low natural isotopic abundance of $^{13}$C.

As shown in figure~\ref{f:secondaries_range}, the particle content of muon-induced showers is, as expected~\cite{groom}, dominated by $\gamma$-rays and electrons.  Around 25~MeV, at the giant dipole resonance of $^{12}$C where the cross section for $\gamma$-rays is quite large, the cumulative range of $\gamma$-rays is two to three orders of magnitude larger than the corresponding values for neutrons, and four to five orders of magnitude larger than for other hadrons.  As a consequence, the dominant process for the production of $^{11}$C nuclides is the ($\gamma$,n) exchange reaction, accounting for $\sim$60\% of the total production rate, even if the hadronic cross sections for the $^{11}$C production are up to a factor~10 larger than the peak value for $\gamma$-rays.  Electrons are not as effective as $\gamma$-rays because their cross section is lower by a factor $\alpha$.  Other hadronic channels with a neutron in the final state and the muon-induced photodisintegration, also carrying a neutron, account for an additional 35\% of the total production rate.  The rate of $^{11}$C production in the two invisible channels, corresponding to the (p,d) and the ($\pi^+$,$\pi^0$+p) exhange reactions, accounts for only about 5\% of the total production rate. 

Deutsch's idea of eliminating the $^{11}$C events by looking at the three-fold coincidence turns out to be still valid even in presence of invisible channels: only 1 out of 20~$^{11}$C nuclides is produced without a neutron in the final state.  In section~\ref{s:reduction} we will quantify the effectiveness of the $^{11}$C background reduction technique based on the double muon+neutron tagging.

\section{Comparison with previous estimates}
\label{s:comparison}

O'Connell and Schima~\cite{oconnell} calculated the production rate of several radioactive nuclides in carbon, oxygen, and argon targets, at sea level and at the depth of KamLAND, only for the photoproduction induced by the virtual photons associated with the muons.  For the $^{11}$C nuclides, they took into consideration only the channel $^{12}$C($\mu$,$\mu$+n)$^{11}$C which results, in our estimate, in a 5\% contribution to the total production rate.  They estimated a production rate of $^{11}$C of 15~events per day in 100 tons of carbon at the KamLAND depth, or, scaling with the muon flux quoted in table~\ref{t:muons}, about $\rm 6.0 \times 10^{-4} \ nuclides/\mu/m$ track in organic liquid scintillators.  Their result is to be compared with our estimate for the same channel reported in table~\ref{t:prod-rates}, which amounts to $\rm 2.4 \times 10^{-4} \ nuclides/\mu/m$.  The estimate of O'Connell and Schima is about 2.5~times higher than our estimate.  The reason for the discrepancy lies in the value chosen for $\sigma_{-1}$, the inverse energy weighted moment of the photodisintegration cross section.  The value of $\sigma_{-1}$ quoted in~\cite{oconnell} is 4.5~mb, while the value we calculated from the cross sections in use in this paper is 1.74~mb.  It is worth noting that work cited as the source for the photoneutron cross sections in the paper of O'Connell and Schima, the {\it Atlas of photoneutron cross sections obtained with monoenergetic photons}~\cite{berman}, offers two values for $\sigma_{-1}$ extrapolated from three different experiments, whose average is 1.65~mb.  This is in excellent agreement with our calculation and is about a factor~2.5 lower than the value quoted in~\cite{oconnell}.

\section{$^{11}$C reduction techniques}
\label{s:reduction}

As shown in the previous section, at least one neutron is produced in association with 95\% of the $^{11}$C nuclides.  This fraction of the $^{11}$C background can be lowered with the muon+neutron tagging.  The remaining 5\% cannot be reduced with this technique.

In order to suppress the $^{11}$C background, one needs to identify the position and time of each neutron created by a muon-induced shower and then captured on protons in the scintillator.  Neutron capture on protons results in the emission of a distinctive 2.2~MeV $\gamma$-ray.  In hydrocarbons, neutrons can also be captured on $^{12}$C resulting in $\gamma$-rays of combined energy 4.9~MeV~\cite{pgaa}.  The cross section for the capture on $^{12}$C is $\sim$1\%~\cite{neutron_capture} of the cross section for capture on protons.  Once the neutron capture time and position are known, one needs to apply a cut in space and time around every capture point.  The events within a time t from the double muon+neutron coincidence and inside a sphere of radius r from the neutron capture point are rejected.  This technique, originally suggested in~\cite{deutsch1}, has recently been succesfully applied in the 4-ton prototype Counting Test Facility of Borexino~\cite{franco}.

\begin{figure}[t]
\includegraphics[width=3.4in]{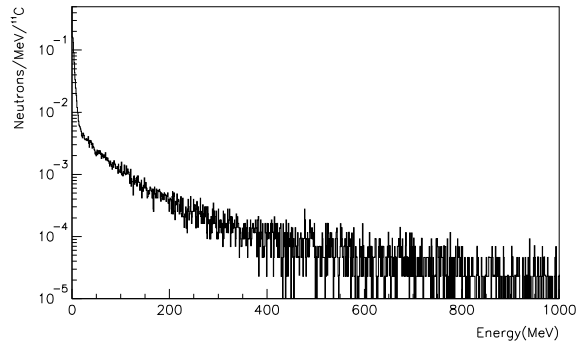}
\caption{\label{f:neutrons_energy} Energy distribution for neutrons produced in association with $^{11}$C nuclides.}
\end{figure}

\begin{figure}[t]
\includegraphics[width=3.4in]{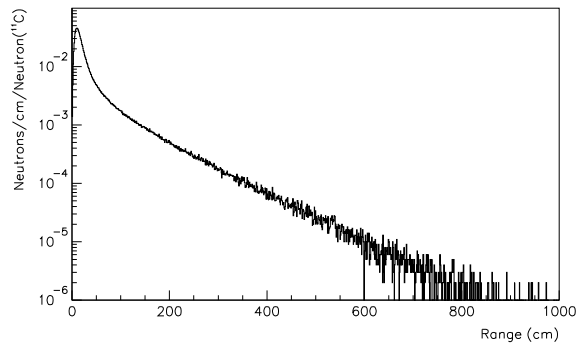}
\caption{\label{f:neutrons_range} Range of neutrons produced in association with $^{11}$C nuclides.}
\end{figure}

The length of time t for which events are rejected should be set to a few times the mean life of $^{11}$C.  Note that the information carried by the neutron capture does not tell us about the position of the $^{11}$C birthplace.  Therefore, it is important to set the radius r of the spherical cut to a few times the average neutron range (Note: the spatial resolution of organic liquid scintillator detectors is about 10~cm~\cite{borex} for the 2.2~MeV $\gamma$-rays from neutron capture and can be neglected with respect to the average neutron range).
For this reason, we calculated with Monte Carlo methods the energy distribution of the neutrons produced in association with $^{11}$C nuclides for Borexino, KamLAND and at SNOLab.  The calculation was performed using the relative weights of the $^{11}$C production in the different channels as determined for muons of 320 GeV.  The procedure is correct for all the three locations since the relative weights $^{11}$C production in the different channels at 285 and 350 GeV are within 3\% of the values at 320 GeV (see table~\ref{t:prod-rates}).  We used the resulting distribution, shown in figure~\ref{f:neutrons_energy}, to source neutrons into FLUKA, calculating the distribution of the range of neutrons produced in association with $^{11}$C, shown in figure~\ref{f:neutrons_range}.

The average energy of neutrons produced in association with $^{11}$C is much lower than the average energy of all neutrons produced in muon-induced showers: this is due to the dominant production mechanism being the photoneutron reaction at the giant dipole resonance of $^{12}$C at 23~MeV.  Correspondingly, the average range of neutrons associated with $^{11}$C production is also much lower than the average range of all the neutrons produced in the shower.

In our calculation we assumed that the $^{11}$C nuclides displacement between the point where it is created and the point where it decays is negligible with respect to the range of neutrons created in association with $^{11}$C nuclide.  For that to happen, the convective motion of the scintillator has to be sufficiently slow.  This can be achieved, for example, by maintaining a small temperature gradient pointing upward everywhere in the detector.  The average range of neutrons created in association with $^{11}$C nuclides, whose distribution is shown in figure~\ref{f:neutrons_range}, is 44~cm.  KamLAND data~\cite{shirai} show that the measured average displacement of the diffusive $^{222}$Rn over its 5.5~days mean life is less than 1~meter.  Therefore the assumption that the $^{11}$C nuclides displacement over their 30~minutes mean life can be kept small with respect to the neutrons range seems fully justified.

Due to the presence of the (n,2n) exchange reaction that yields two neutrons in the final state, an average of 1.14 neutrons are created in interactions producing $^{11}$C nuclides.  Those neutrons, if sufficiently energetic, can also trigger nuclear reactions knocking off other neutrons: our calculation indicates that an average number of about 1.2 neutrons are captured in the scintillator for each neutron produced in a $^{11}$C-forming reaction.

The efficiency $\epsilon$ of rejecting $^{11}$C events tagged with the muon+neutron coincidence is equal to the combined efficiencies for the cut in space, $\zeta$(r), and in time, $\eta$(t):

\begin{equation}
\label{e:eff}
\rm \epsilon = \zeta(r) \cdot \eta(t)
\end{equation}
where $\eta(t)$ is the efficiency for the rejection of $^{11}$C events when a cut in time for a time span equal to t is applied around the neutron capture point.  Given the expected decay distribution of $^{11}$C, we obtain:

\begin{equation}
\label{e:eta}
\rm \eta(t) = 1 - e^{-t/\tau}
\end{equation}
where $\tau$ is the mean life of $^{11}$C, 30 minutes.

In equation~\ref{e:eff}, $\zeta(r)$ is the rejection efficiency for $^{11}$C when a cut in space is applied around the neutron capture point, corresponding to a sphere of radius r centered around the capture point position.  Given the distribution n(s) for the range of neutrons shown in figure~\ref{f:neutrons_range}, the efficiency for the rejection is equal to:
\begin{equation}
\label{e:zeta}
\rm \zeta(r) = \int_0^r n(s) ds\left/ \int_0^\infty n(s) ds \right.
\end{equation}

In order to quantify the effectiveness of the background reduction technique we introduce a figure of merit R, independent from the experiments and their locations.  R~is defined as the ratio of the pep+CNO neutrino signal rate (S) to the residual background rate (B) from $^{11}$C after suppression of all $^{11}$C events identified through the the muon+neutron coincidence (both S and B are computed in the pep energy window [0.8-1.3~MeV]).  Note that the figure of merit R accounts only for the remaining background from $^{11}$C, but other possible and independent sources of background are neglected in its definition.  The general expression for R is:

\begin{equation}
\label{e:merit}
\rm R = S/B = \frac{S/B_0}{F + \left( 1- F \right) \left( 1-\zeta\eta \right)} 
\end{equation}
where $\rm F \simeq 0.05$ is the fraction of $^{11}$C production rate in invisible channels; S (signal rate) was discussed in section~\ref{s:introduction} and B$_0$ (raw $^{11}$C background rate, experiment- and location-dependent) was discussed in section~\ref{s:rates} and in table~\ref{t:c11-extr}.  Formul\ae~\ref{e:eff} and~\ref{e:merit} show that there is a one-to-one correspondance between the figure of merit R and the combined rejection efficiency $\epsilon$.

Using equation~\ref{e:merit}, the signal rate reported in section~\ref{s:introduction}, and background rates from table~\ref{t:c11-extr}, we estimate that the optimal signal to background ratio achievable in Borexino (KamLAND) is 8 (1) in case all of the $^{11}$C associated with a neutron are successfully tagged.  The optimal limit cannot however be reached due to dead mass-time limitations, as shown below.

We define the dead mass-time fraction D as the fraction of (mass $\times$ time) data taking lost to the space and time cuts around a neutron capture event.  Treating $\eta$ and $\zeta$ as independent variables, t and r can be derived by inverting equations~\ref{e:eta} and~\ref{e:zeta} and become functions of $\eta$ and $\zeta$ respectively.  We can then compute the dead mass-time fraction D corresponding to the chosen values of the cuts $\zeta$ and $\eta$ as:

\begin{equation}
\rm D = \rm 1 - e^{-\frac{4}{3} \pi \rho r^3 t N}
\end{equation}
where $\rho$ is the scintillator density, and N is the neutron rate per unit mass in the detector.  The expected neutron rates for the experiments taken into consideration are summarized in table~\ref{t:muons}.

\begin{figure}[t!]
\includegraphics[width=3.4in]{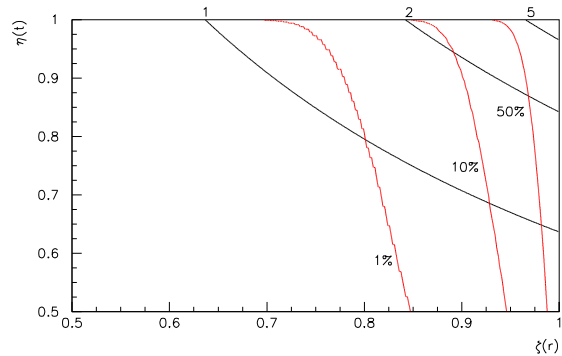}
\caption{\label{f:dead_masstime}Correlation between the signal to reduced background ratio (R) and the dead mass-time fraction for Borexino, as a function of spatial ($\zeta$) and time ($\eta$) efficiencies of $^{11}$C events rejection.  The only background considered in the pep energy window [0.8-1.3~MeV] is the $^{11}$C background surviving the cuts described in the text.  Isocontours for the ratio R (for values $\rm R = 1, 2, 5$) and for the dead mass-time fraction D (for values $\rm D = 1\%, 10\%, 50\%$) are shown in the graph.}
\end{figure}

Figure~\ref{f:dead_masstime} shows the correlation between the signal to reduced background ratio R and the dead mass-time fraction D for Borexino, as a function of spatial~($\zeta$) and time~($\eta$) efficiencies for $^{11}$C events rejection.
In table~\ref{tableratio} we show the optimal values of the dead mass-time fraction D for the three experiments for fixed and given values of the ratio R between signal and reduced $^{11}$C background.  The values are calculated by minimizing the dead mass-time fraction D while keeping the product of the two efficiencies $\zeta$, $\eta$ constrained to the value of $\epsilon$ corresponding to the chosen value of R.

\begin{table}[t!]
\caption{\label{tableratio} Optimal values of the dead mass-time fraction (D) for KamLAND, Borexino and SNO as a function of the figure of merit R, defined as the ratio between the signal rate (S) of pep and CNO neutrinos (BP04 model for MSW-LMA scenario~\cite{bp04}) and background rate (B) from $^{11}$C after the reduction with the technique outlined in the text.  In Borexino the ratio between signal rate and raw background before any $^{11}$C rejection (B$_0$) is 0.4; in KamLAND, S/B$_0$ is 0.05; for a detector at SNOLab, S/B$_0$ would be 36.}
\begin{ruledtabular}
\begin{tabular}{dddd}
		\multicolumn{1}{c}{S/B$_0$}
		& 0.05
		& 0.4
		& 36\\
\hline\hline
		&\multicolumn{1}{c}{KamLAND}
		&\multicolumn{1}{c}{Borexino}
		&\multicolumn{1}{c}{@ SNOLab} \\
		\multicolumn{1}{c}{R = S/B}
		&\multicolumn{1}{c}{D [\%]}
		&\multicolumn{1}{c}{D [\%]}
		&\multicolumn{1}{c}{D [\%]} \\
\hline
0.1		&0.4			&			&\\
0.2		&11.6		&			&\\
0.3		&50.6		&			&\\
0.4		&87.4		&<0.1		&\\
0.5		&98.8		&<0.1		&\\
0.8		&>99.9		&0.1			&\\
1		&			&0.3			&\\
2		&			&6.7			&\\
3		&			&27.8		&\\
4		&			&58.3		&\\
5		&			&85.3		&\\
8		&			&>99.9		&\\
100		&			&			&<0.1\\
500		&			&			&2.6\\
\end{tabular}
\end{ruledtabular}
\end{table}

Further improvements of these figures might be achieved by the experiments by optimizing the set of cuts used to tag $^{11}$C events with the double muon and neutron coincidence according to each detector's capabilities and performance.  For example, the accurate reconstruction of the track of the through-going muons might help reducing the dead mass-time by modifying the topology of the spatial cut proposed in this study, such as limiting the volume of the region excluded by the cuts to an intersection of a cylinder around the muon track and of a sphere centered around the neutron capture point, as proposed in~\cite{franco}.

The technique offers as a by-product the possibility of determining the total number of $^{11}$C decays in the detector.  The residual background B can then be statistically subtracted from the spectrum of the reduced data set.  Let T be the total rate in the pep window after applying the muon+neutron tagging technique.  Under the assumption that the internal and the external background rates are negligible with respect to the residual background from $^{11}$C and to the signal rates, $\rm S = T - B$.  A lower bound on the statistical error associated with the signal rate S extrapolated with the statistical subtraction of the energy spectra is obtained by propagating the errors in the formula just introduced:

\begin{eqnarray}
\label{e:stat-err}
\rm \frac{\delta S}{S} & = & \rm \frac{\sqrt{N_T+N_B}}{N_S} 
= \frac{\sqrt{N_S+2N_B}}{N_S}
\nonumber \\
 & = & \rm \sqrt{\frac{1+2/R}{S M (1-D) t_0}} 
\end{eqnarray}
where $\rm N_T$, $\rm N_S$, and $\rm N_B$ are, respectively, the total number of events recorded, the total number of signal events and the total number of background events (all after application of the cuts to reduce the background from $^{11}$C), t$_0$ is the data taking time and M is the fiducial mass for pep and CNO neutrinos observation. 
The formula above can be used to obtain information concerning the statistical accuracy of the measurement of the signal rate.  Borexino (KamLAND) with a 70-ton (300-ton) fiducial mass would achieve a statistical accuracy of 2.9\% (3.4\%) in 5 years.

\section{Concluding remarks}
\label{s:conclusions}

In this paper we presented a study of the production mechanism of $^{11}$C nuclides in muon-induced showers.  We identified the nuclear reactions relevant for the production of the nuclide in muon-induced showers in organic liquid scintillators.  We performed an {\it ab initio} calculation of the production rates for each channel, then compared the calculated total production rate with available experimental data, obtaining a good agreement.  We estimated that for 95\% of the $^{11}$C nuclides produced at least one neutron is emitted.

A possible experiment located at SNOLab has a very low muon flux and hence a $^{11}$C production rate which is negligible when compared with the expected rate from pep and CNO neutrinos.  On the other hand, for both Borexino and KamLAND, cosmogenic $^{11}$C is a significant background for the detection of pep and CNO solar neutrinos.

We discussed a reduction technique for the $^{11}$C events, based on one-by-one identification through the coincidence between a parent muon and the resulting neutron capture.  We estimated that both Borexino and KamLAND could use the technique to increase the original signal (pep+CNO neutrinos) to background ($^{11}$C) ratio by a significant factor.  Borexino can improve from a signal/background ratio of 0.4 to one of 2 (4) while losing 7\% (58\%) of the data to dead mass-time.  KamLAND can improve its signal/background ratio from 0.05 to 0.2 (0.3) while losing 12\% (51\%) of data to dead mass-time.

The residual $^{11}$C background can be statistically subtracted to determine the signal rate from neutrinos.  We presented a formula providing a crude estimate of the statistical error associated with the signal rate determined with this procedure.  We estimated that both Borexino and KamLAND could measure the combined rate from pep and CNO neutrinos in the [0.8-1.3~MeV] window down to a 3\% statistical accuracy in five years, provided that internal and external background rates are kept within figures negligible with respect to the signal rate.

\section{Acknowledgements}
We thank F.~Calaprice for useful discussions, for the suggestion to investigate all production channels, and for drawing our attention to the (p,d) exchange reaction.  We thank C.~Pe{\~n}a-Garay and F.~Vissani for reading the manuscript and for useful comments.  We thank G.~Battistoni for help in getting started with FLUKA simulations.  The work of C.G. and A.P. was supported in part by the U.S. National Science Foundation under grant \mbox{PHY-0201141}.  L.C. acknowledges the support of the U.S. National Science Foundation under grant \mbox{PHY-0107417}.  C.G. acknowledges hospitality from the Dipartimento di Fisica of the Universit\`a degli Studi di Milano during part of this work.

In memory of M.~Deutsch, for inspiring discussions on the subject of cosmic ray-induced interactions and for providing the founding idea behind this study.

\end{document}